\documentclass[12pt,twoside]{article}
\usepackage{fleqn,espcrc1}

\usepackage{graphicx}

\usepackage[figuresright]{rotating}

\newcommand{\AmS}{{\protect\the\textfont2
  A\kern-.1667em\lower.5ex\hbox{M}\kern-.125emS}}
\title{Initial singlet and triplet spin state contributions 
to~$\overrightarrow p \overrightarrow p \rightarrow pp\pi^0$}
       
\begin{document}
\maketitle
\noindent 
P. Th\"orngren Engblom$^a$\footnote{Present address:Department of Radiation Sciences, 
Box 535, S-75121 Uppsala, Sweden.\\
Email: pia.thorngren@tsl.uu.se}, H.O. Meyer$^a$, J.T. Balewski$^a$, W.W. Daehnick$^b$, J. Doskow$^a$,\\
W. Haeberli$^c$, B. Lorentz$^c$, P.V. Pancella$^d$, R.E. Pollock$^a$, B. von Przewoski$^a$,\\
F. Rathmann$^e$,T. Rinckel$^a$, Swapan K. Saha$^b$, B. Schwartz$^c$, A. Wellinghausen$^a$, T. Wise$^c$

\vspace{4mm}
\noindent{$^a$Indiana University Cyclotron Facility, Indiana University, Bloomington, IN 47405, U.S.A.}\\
\noindent{$^b$Department of Physics, University of Pittsburgh, Pittsburgh, PA 15260, U.S.A.}\\
\noindent{$^c$Department of Physics, University of Wisconsin-Madison, Madison, WI 53706, U.S.A.}\\
\noindent{$^d$Department of Physics, Western Michigan University, Kalamazoo, MI 49008, U.S.A.}\\
\noindent{$^e$Friedrich-Alexander Universit\"at, Erwin-Rommel Str. 1, 91508 Erlangen, Germany.}
\vspace{8mm}
\begin{abstract}
The PINTEX\footnote{http://www.iucf.indiana.edu/$\sim$pintex/} facility at the 
IUCF Cooler ring, dedicated to the study of spin dependence in nucleon-nucleon
interactions, has been used to measure polarization observables of the reaction
$\overrightarrow p \overrightarrow p \rightarrow p p \pi^0$ at beam energies
between 325 and 400 MeV. The stored, polarized proton beam had spin projections
both in the longitudinal and the transverse directions with respect to the beam 
momentum. 
We report here on the measurements of the relative transverse and longitudinal
spin-dependent cross sections\footnote{Defined as 
$\Delta\sigma_{T(L)}=[\sigma (\uparrow\downarrow)+ \sigma (\downarrow\uparrow)] - 
[\sigma (\uparrow\uparrow + \sigma (\downarrow\downarrow)]$
where the arrows denote parallel and antiparallel beam/target spin combinations,
either transversely (T) or longitudinally (L) polarized.},
$\Delta \sigma _T/\sigma _{tot}$ and $\Delta \sigma _L/\sigma _{tot}$, and how from 
these observables the initial spin singlet and triplet cross sections 
are obtained. Considering angular momentum states less than or equal to one,
the contribution of the Ps partial waves to the cross section can be extracted.
\end{abstract}
\section{INTRODUCTION}
For reactions with two incoming spin 1/2 particles one can define 
the initial state cross sections $^{2S+1}\sigma _{m_s}$ 
where $S$ denotes the combined spin of the two particles and $m_s$ its 
projection. The three possible initial spin state contributions 
$^{1}\sigma _{0}$, $^{3}\sigma _{0}$ and  $^{3}\sigma _{1}$
are related to 
$\Delta \sigma _T$ and $\Delta \sigma _L$
by isospin, parity and angular momentum conservation
\begin{equation}
\frac{^1\sigma_{0}}{\sigma _{tot}} =
1 + \frac{\Delta\sigma_T}{\sigma _{tot}} + \frac{1}{2}
\frac{\Delta\sigma_L}{\sigma _{tot}}
\end{equation}
\begin{equation}
\frac{^3\sigma_{0}}{\sigma _{tot}} =
1 - \frac{\Delta\sigma_T}{\sigma _{tot}} + \frac{1}{2}
\frac{\Delta\sigma_L}{\sigma _{tot}}
\end{equation}
\begin{equation}
\frac{^3\sigma_{1}}{\sigma _{tot}} =
1 -  \frac{1}{2}\frac{\Delta\sigma_L}{\sigma _{tot}}
\end{equation}


\clearpage
Near threshold only a limited number of partial waves are important
and considering the allowed transitions to final states\footnote{
$L_{NN}$ is the angular momentum between the two
nucleons and $l_{\pi}$ is the angular momentum of the pion with respect to
the nucleon pair.} with $L_{NN}\leq 1$
and $l_{\pi}\leq 1$, i.e. assuming
$\sigma _{tot} \equiv \sigma_{Ss}+\sigma_{Ps}+\sigma_{Pp}$, we obtain the 
following relations\footnote{$Sp$ is 
forbidden by parity and angular momentum conservation} 
\begin{equation}
^1\sigma_{0} = 4 \sigma_{Ps}
\end{equation}
\begin{equation}
^3\sigma_{0} = 4 (\sigma_{Ss} + \hat{\sigma}_{Pp})
\end{equation}
\begin{equation}
^3\sigma_{1} = 2 (\sigma_{Pp} - \hat{\sigma}_{Pp})
\end{equation}
where $\hat{\sigma}_{Pp}$ corresponds to $S=1$, $m_s=0$ initial states that
contribute to $Pp$ final states but not to $\sigma_{tot}$.
To which extent the Ps partial waves contribute to $\sigma_{tot}$ 
can then be obtained in a model-free way\cite{HOM99} using equ. 1-6
\begin{equation}
\frac{\sigma_{Ps}}{\sigma_{tot}} =
\frac{1}{4}(1 + \frac{\Delta \sigma_T}{\sigma_{tot}} + 
\frac{1}{2}\frac{\Delta \sigma_L}{\sigma_{tot}})
\end{equation}
Theoretical models for pion production near threshold can thus
be tested in an unprecedented manner using measurements of spin 
observables of the reaction 
$\overrightarrow p \overrightarrow p \rightarrow p p \pi^0$\cite{HOM98C}.
The results for the spin-correlation coefficient combinations $A_{\Delta}=A_{xx}-A_{yy}$,
and $A_{\Sigma}=A_{xx}+A_{yy}$ (the latter equivalent to $-\Delta\sigma_T/\sigma_{tot}$) and the 
analyzing power, $A_y$, which only needed transverse beam and target polarization,
were published recently\cite{HOM98} and the publication of the 
the data from the experiment using also longitudinal polarization 
components is in progress\cite{HOM99}.
The simultaneous measurements of spin correlation coefficients in 
$\overrightarrow p \overrightarrow p \rightarrow p n \pi^+$
will further enhance the usefulness of this experiment for
resolving the theoretical issues of $NN \rightarrow NN \pi$ reactions\cite{DAEHNICK99}.

Section 2 contains a brief description of the experimental method which
is described in detail elsewhere\cite{PTE98C,NIM99}.
In section 3 we discuss the preliminary results of the measurements using 
polarized beam with longitudinal components, and the possible implications.

\section{THE EXPERIMENT}
The vertically polarized beam was injected into the IUCF Cooler ring
at 197 MeV and then accelerated to 325, 350, 375 or 400 MeV. The longitudinal 
component of the beam polarization was achieved using one superconducting 
solenoid and the three existing solenoids in the cooling section operated in
a non-compensating mode. At the injection energy the polarization at 
the target was almost entirely longitudinal but since the solenoid fields 
remained constant, after acceleration only partially longitudinal 
beam polarization could be achieved.
The major components were along and vertical to the beam momentum and one 
minor component was sideways. The beam spin direction was reversed at each 
new injection. 

The target was polarized atomic hydrogen in a gas
storage cell fed by an atomic 
beam source\cite{WISE93}. The target thickness was of the order of 
$10^{13}$ atoms$\rm /cm^2$. The spin direction was changed
every two seconds, along the $\pm x$, $\pm y$ and $\pm z$ axes, by switching
weak guide fields ($\sim 0.3-0.6$ mT), supplied by Helmholtz-like coils 
external to the target chamber. A total of 12 beam/target spin combinations 
were used. 
\begin{figure}[htb]
\begin{minipage}[t]{78mm}
\includegraphics[width=75mm]{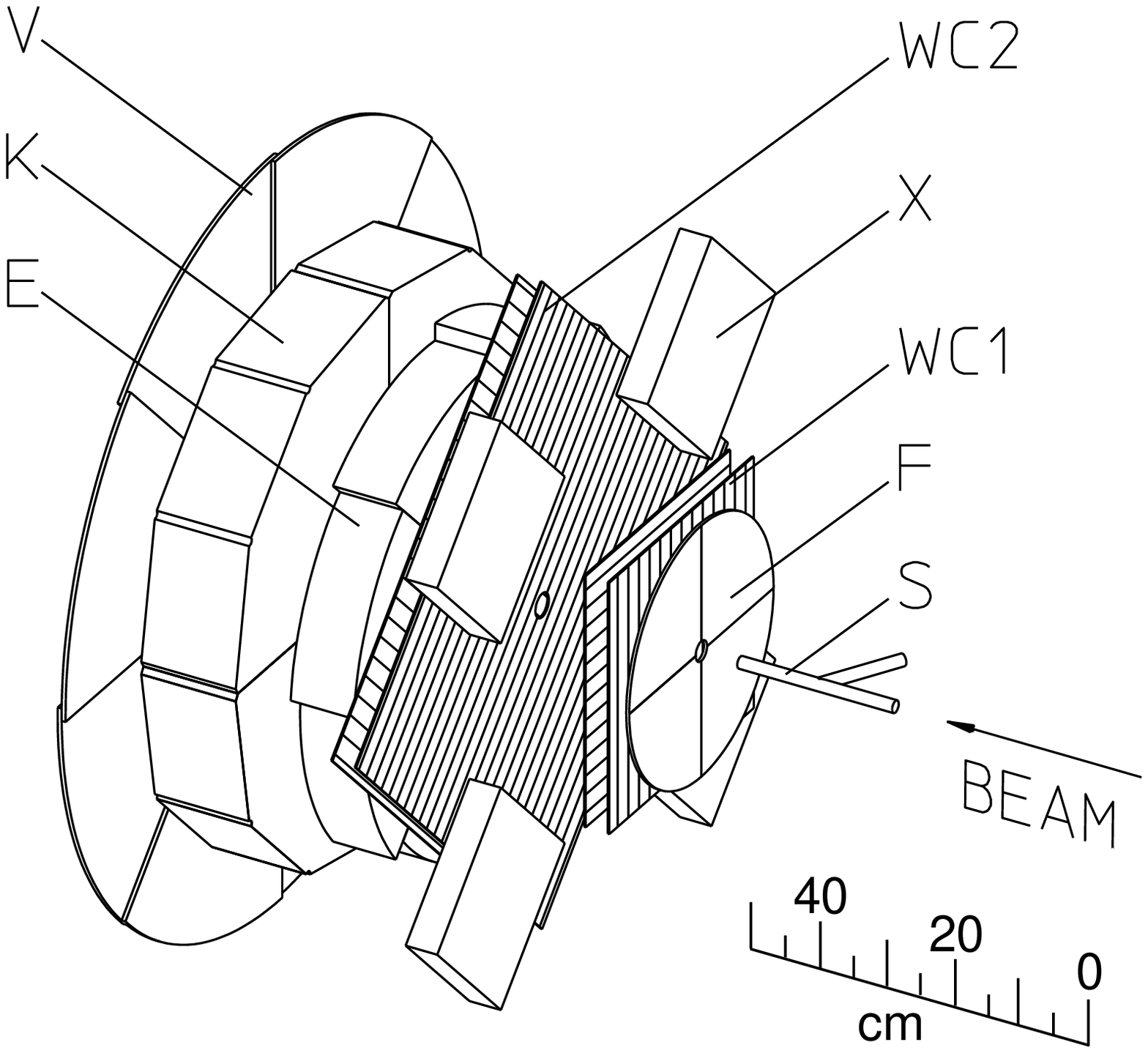}
\caption{The forward detector set-up with an angular coverage of 
$5^{\circ} \leq \theta_{lab} \leq 32^{\circ}$.
The different parts are indicated:
S: storage cell, F: front detector used for time-of-flight measurements, 
E, K: stopping scintillators with eight and four segments, thicknesses
10 and 15 cm respectively, V: veto detector, WC1,2: wire chambers, 
X: four monitor scintillators for measuring $pp$ elastic scattering at 
$\theta_{lab}=45^{\circ}$ and $\phi=\pm 45^{\circ}$ and $\pm 135^{\circ}$.}
\label{fig:TURKEY}
\end{minipage}
\hspace{\fill}
\begin{minipage}[t]{78mm}
\includegraphics[width=75mm]{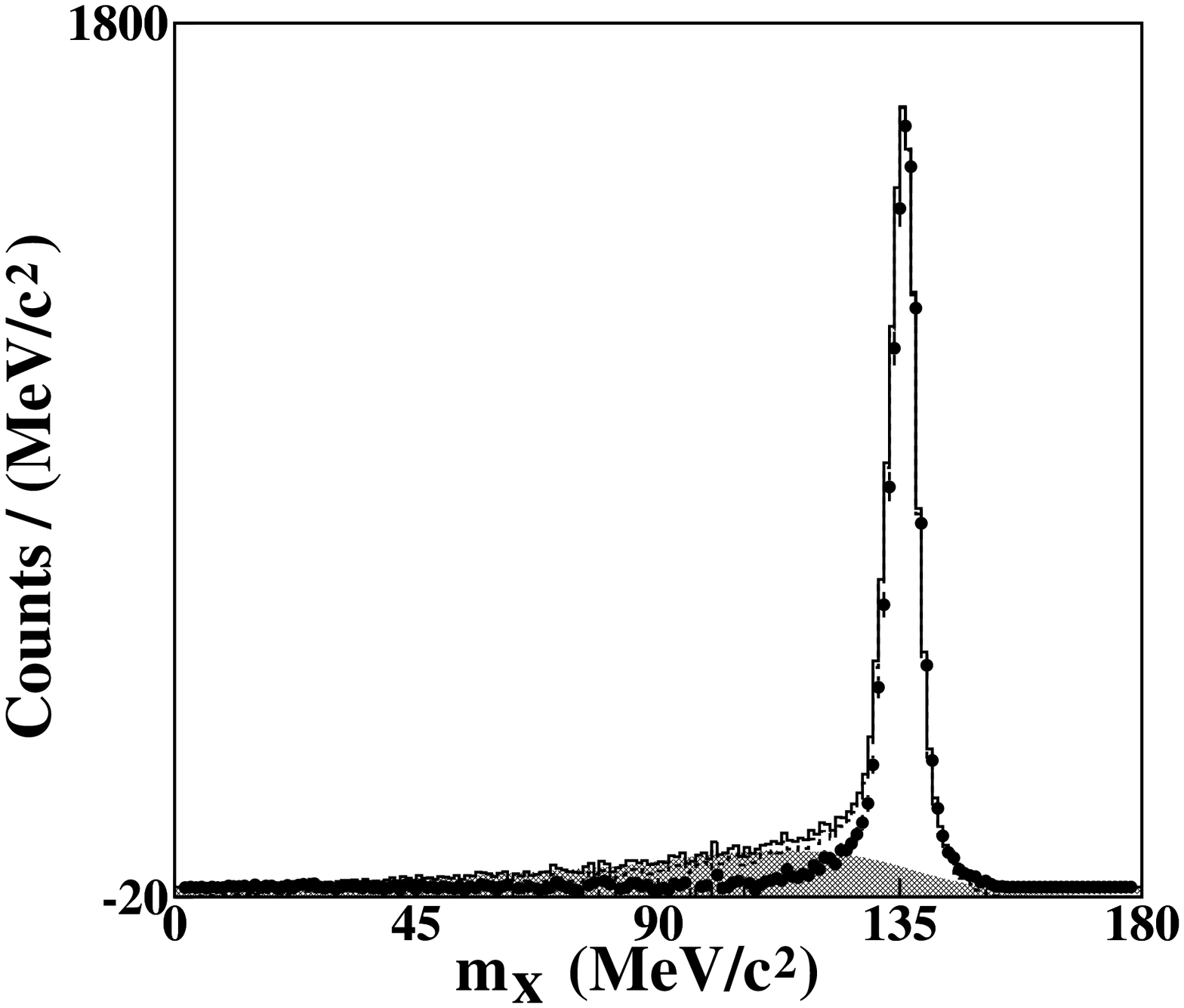}
\caption{The unpolarized distribution of the missing mass 
at 325 MeV beam energy in bins of 1 MeV/c$^2$.
The shaded area indicates the background which has been 
normalized to the area corresponding to invariant masses below 100 MeV/c$^2$.
The filled circles represent the data after the background has 
been subtracted, the statistical errors are shown.
The resolution (FWHM) of the 
background subtracted peak is 7.2 MeV/c$^2$.}
\label{fig:MX}
\end{minipage}
\end{figure}
The four-momenta of the two outgoing protons were measured in a
cylindrically symmetric forward detector stack, see Fig. \ref{fig:TURKEY}, 
which allowed for the missing 
mass, $m_x$, of the third particle to be calculated. In Fig. \ref{fig:MX} 
the $m_x$ distribution is shown for the data set with longitudinal and 
transverse beam spin components summed over all spin states at a beam energy 
of 325 MeV. The shape of the background in the spectrum was well reproduced by 
data taken with nitrogen gas instead of hydrogen in the target.

Full coverage of phase space was achieved with the exception of the 
loss of events when protons go undetected into the beampipe, i.e.
$30-22\%$ of phase space distributed events were lost at 325-400 MeV.
In an investigation of systematic errors the spin observables were studied
as a function of an artificial $\theta_{lab}$ cut-off using an effective-range
expansion for the $pp$ final state interaction\cite{HOM92}. The corrections
applied for the relative spin-dependent cross sections were less
than the statistical errors.

The cross-ratio method\cite{CROSSRATIO} which eliminates asymmetries due to detector
acceptance and spin-dependent luminosities, and division by the 
product\footnote{$\Delta\sigma_{T(L)}/\sigma_{tot} 
= \frac{2}{PQ}\frac{\sqrt{R}-1}{\sqrt{R}+1}$
where $R$ is the ratio of the yields, 
$Y_{\uparrow\downarrow}Y_{\downarrow\uparrow}/
Y_{\uparrow\uparrow}Y_{\downarrow\downarrow}$ and $PQ$ is the product
of the beam and target polarization.} of 
the beam and target polarization projections, $P_yQ_y$ or $P_zQ_z$, give
$\Delta\sigma_T/\sigma_{tot}$ and $\Delta\sigma_L/\sigma_{tot}$ from 
the spinsorted and backgroundsubtracted yields
of the $\pi^0$.
The values of $P_yQ_y$ and $P_zQ_z$ are obtained.
by detecting $pp$ elastic scattering at $\theta_{lab}=45^{\circ}$
simultaneously with the pion production and measuring the $pp$ spin
correlation coefficients $A_{\Delta}$ and $A_{zz}$ which are large
and known with high precision at this angle\cite{HAEBERLI97,RATHMANN98,PRZEWOSKI98,LORENTZ98}.
\section{PRELIMINARY RESULTS AND DISCUSSION}
\begin{figure}[htb]
\includegraphics[width=125mm]{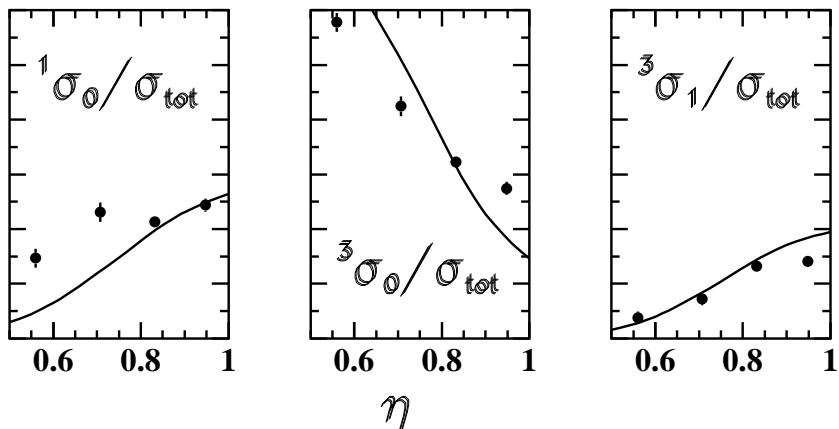}
\caption{The data points are the preliminary results for the initial spin state cross sections. 
The curves are the predictions by the J\"ulich meson-exchange model
(see text). The divisions on the y-axis are arbitrary.}
\label{fig:Sms}
\end{figure}

The analysis of the experiment described here is near completion\cite{HOM99}.
Preliminary results using equ. 1-3 are shown in Fig. \ref{fig:Sms}. Also depicted are
curves representing predictions by a microscopic model calculation,
which is based on direct production and off-shell rescattering diagrams.
Contributions from heavy meson-exchange $(\omega)$ are added and the $\Delta$-isobar
is included explicitly\cite{HANHART98,HANHART99C}. Previous comparison with data 
yielded qualitative agreement for $-A_{\Sigma}$ and $A_y$, 
but disagreed completely with the data on $A_{\Delta}$\cite{PTE98C}. 
Fig.~\ref{fig:Sms} indicates that
$^3\sigma_{1}$, the initial state cross section contributing only to Pp final states, 
contains the amplitudes quantitavely best reproduced by theory so far.

In \cite{HOM99} it is shown that the energy dependence assumed for the Ps
and Pp partial waves, i.e. the proportionality to $\eta^6$ and $\eta^8$ respectively, is
valid for the measured energy range. The magnitude of
the Ps cross section is calculated using equ. 7, it increases from about
18\% to 30\% for bombarding energies in the range from 325 to 400 MeV.
At 310 MeV there are recent reports of a significant Ps contribution to the
total cross section\cite{ZLO99C} and an Sd-Ss interference of a few 
percent\cite{ZLO98}, measuring unpolarized angular distributions.

We conclude that it should be of high priority to proceed with the study of the main inelastic 
channel in $NN$ interactions, both experimentally by means of high statistics
experiments that allow the extraction of angular distributions of spin-dependent 
observables, and theoretically by extending the many existing models concerning the 
Ss contribution to the total cross section to also include higher partial waves.

\section*{Acknowledgements}
We thank Dr. C. Hanhart  for making the theoretical predictions 
of the model of the J\"ulich group available and
one of the authors, P.T.E., thanks Dr. C. Hanhart for useful discussions
and comments.
This work has been supported by the US National Science Foundation under 
Grants PHY95-14566, PHY96-02872, PHY97-22556, and by the US Department 
of energy under Grant DOE-FG02-88ER40438.

\end{document}